\documentclass[3p,times,twocolumn]{elsarticle}

\usepackage{ecrc}

\volume{00}

\firstpage{1}

\journalname{Nuclear Physics B Proceedings Supplement}

\runauth{Shou-hua Zhu}

\jid{nuphbp}

\jnltitlelogo{Nuclear Physics B Proceedings Supplement}

\usepackage{amssymb}
\usepackage{amsmath}
\usepackage{color}
\usepackage{natbib}

\usepackage{hangcaption,fancybox,feynmp}
\usepackage{indentfirst}
\usepackage{graphicx}
\usepackage{epsfig,subfigure,psfrag}
\usepackage{mathrsfs}
\usepackage{slashed}
\usepackage{url} 

\usepackage[figuresright]{rotating}

\begin{document}

\begin{frontmatter}



\dochead{}

\title{Charged Higgs Boson: Tracer of the Physics beyond Standard Model}


\author{Shou-hua Zhu}

\address{$ ^1$ Institute of Theoretical Physics $\&$ State Key Laboratory of
Nuclear Physics and Technology,\\
Peking University, Beijing 100871, China \\
$ ^2$ Collaborative Innovation Center of Quantum Matter, Beijing, China \\
$ ^3$ Center for High Energy Physics, Peking University,
Beijing 100871, China}

\begin{abstract}

Charged Higgs boson can exist in many physics beyond the standard models (BSM) and it is the obvious BSM signal. We briefly describe why the 125GeV scalar discovered at the LHC must have (heavy) companion: the charged Higgs boson, in a new paradigm. We then focus on the charged Higgs phenomenology, especially on how to measure $\tan\beta$ precisely utilizing the top quark polarization information.

\end{abstract}


\end{frontmatter}


\section{Introduction}

In July 2012, the new scalar (dubbed as H(125) in this paper) was discovered by ATLAS and CMS of the LHC, which surprised many theorists including me. Furthermore the subsequent measurements are
still consistent with the predictions of the standard model (SM). Though the physics beyond the SM (BSM) has strong motivations \cite{Zhu:2014hda}, the pursuit of it needs some courage and a little bit of luck.
During the ICHEP2014, I can feel the spirit of Don Quijote, which is best described by a song called "The impossible dream":
"To dream the impossible dream/ To fight the unbeatable foe/ To bear with unbearable sorrow/ To run where the brave dare not go/ \ldots
/And the world will be better for this/ That one man scorned and covered with scars/ Still strove with his last ounce of courage/ To reach the unreachable star".

Now that the neutral scalar H(125) was discovered, one may wonder whether there are more scalars to be discovered, especially the charged Higgs boson. It is the obvious BSM
signature. In the SM, one doublet is enough to generate the gauge boson mass and fermion mass, at the same time to induce
the flavor changing interactions with the right magnitude. It seems that no extra scalars are needed. Therefore in second Sec. II, we will briefly present the motivation
for the charged Higgs boson and in Sec. III, we discuss the different top quark polarization in charged Higgs boson decay and in associated production, In Sec. IV, we focus on how to suppress the backgrounds assuming the charged Higgs boson decaying into top plus bottom. In Sec V, we study
how to measure $\tan\beta$ utilizing the top polarization, especially for the intermediate value which is hard to measure using the cross section information, and last section
contains our conclusion and discussion.

\section{Motivation for charged Higgs boson: A possible new paradigm}

Many BSM require more scalar sectors for various reasons. For example, the minimum supersymmetric standard model (MSSM) requires at least two Higgs doublets.

\begin{figure}[!tbh]
  \centering
   \includegraphics[width=\linewidth]{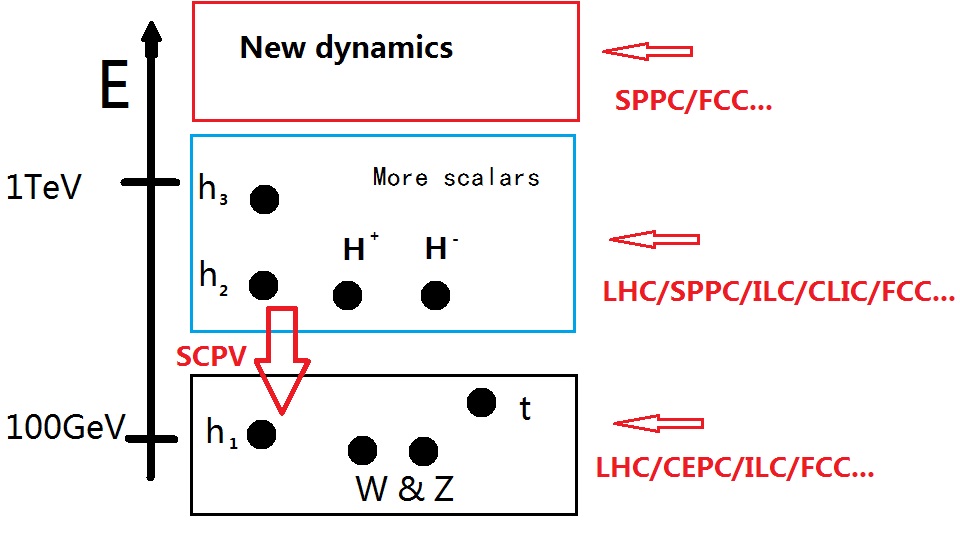}
      \caption[]{ Schematic diagram for a new paradigm. The light H(125) is due to the approximate CP symmetry, and there are other heavier scalars which need to
      be understood. The new dynamics on top of scalar sector is not known yet, which are responsible for the complicated scalar sector and the flavor structure. }
         \label{anp}
  \end{figure}

In the last two years,
we begin to realize a possible new paradigm \cite{Zhu:2014hda} and the schematic diagram of which is shown in Fig. \ref{anp}. This conjectured new paradigm is based on the assumption that
correlation between the lightness of H(125) and the smallness of CP-violation under the framework of spontaneous CP violation (SCPV) \cite{Zhu:2012yv,Hu:2013cda,Mao:2014oya}. The SCPV was firstly proposed by T.D. Lee
in 1973 \cite{Lee}, the original purpose was to find the origin of CP violation. We extended the scope of SCPV and utilized it to bridge the Higgs mass and CPV. In the Lee model, in the $t_\beta s_\xi \rightarrow 0$ (shorthand for $\tan\beta \sin\xi$) limit, namely CP violation vanishing limit, the measures of CP violation (Jarskog variable \cite{Jarskog} and K \cite{2HDM,K}) and lightest Higgs mass can be expressed
as
\begin{equation}
J \propto t_{\beta}s_{\xi}; \ \ \  K\propto t_{\beta}s_{\xi}; \ \ \ m^2_h \propto v^2t^2_{\beta}s^2_{\xi}
\end{equation}
According to such behavior, we propose that the lightness of the Higgs
boson and the smallness of CP-violation effects could be correlated through small $t_{\beta}s_{\xi}$.

Even though there is connection between the lightness of H(125) and small CPV, what implications will it bring to us?
In the past studies, one tried to account for the lightness of the H(125). In the new
paradigm, the mass of H(125) is due to the  approximate CP symmetry and the extra scalar bosons can be much heavier, e.g. at O(TeV) even in
the strong-coupled regime. On top of the scalar sector(s), there is unknown new dynamics which are responsible
for the scalar sectors. This
opens a new approach to understand the electro-weak symmetry breaking and the origin of CP violation.

The generic feature of this new paradigm is summarized as
\begin{itemize}
\item H(125) is not SM-like, and H(125) must be the CP mixing state. This feature is different with other models, in which the SM limit
can be reached in a way or the other.
\item There are usually extra CP violation in scalar sector, besides CKM matrix, though they arise both from the complex vacuum. The CP violation measurements are
important for high and low energy experiments.
\item There are other heavy neutral and charged Higgs bosons.
\item The scale of the new mechanism for the complicated scalar sectors is higher.
\end{itemize}

For the optimistic case (from the point view of experimentalists), the extra new scalars can be not so heavy.
For this case, LHC has the good opportunity to discover them. Therefore, it has good motivation to explore
the discovery potential and capacity to measure its properties for charged Higgs boson.

\section{ Charged Higgs boson and top polarization}

As we have discussed, the charged Higgs boson is quite common in many BSM. If the charged Higgs boson is lighter enough, it can be produced as the
decay product of top quark and charged Higgs boson can subsequently decays into tau plus neutrino. For heavy charged Higgs boson, the main production
channel is  $b g \rightarrow H^+ t^-$. For example, we proposed to search supersymmetric Higgs boson via this channel more than a decade ago \cite{Huang:1998vu} and calculated
the complete QCD correction to this process subsequently \cite{Zhu:2001nt}. In fact there are numerous studies on charged Higgs boson, we refer the interesting reader
to Ref. \cite{Wang2014}.

In this paper, we will focus on how to utilize the top polarization to study the charged Higgs boson  \cite{Cao:2013ud}.
In our recent work~\cite{Cao:2013ud}, we studied this subject in the type-II 2HDM as the benchmark model,
where one Higgs doublet couples to up-type fermions and the other Higgs doublet couples to down-type fermions.
In general, charged Higgs boson can couple with top quark in various ways.

In type-II 2HDM, the coupling among charged Higgs boson and quarks can be written as
\begin{equation}
g_{H^-\bar{d}u}=\frac{g}{\sqrt{2}m_W}(m_d\tan{\beta}P_L+m_u\cot{\beta}P_R),
\label{feynman_tbh}
\end{equation}
where $P_{L/R}=(1\mp\gamma_5)/2$ is the chirality projector.
Here $\tan{\beta}$ is the crucial parameter of 2HDM which is defined as the ratio between the vacuum expectation values of two Higgs doublet.
As indicated in Eq.\ref{feynman_tbh}, the coupling strength is proportional to the mass of corresponding fermions.
Here we will consider the most massive third generation fermions, namely top and bottom. Thanks to the heavy masses, the top quark decays promptly
so that the chirality information of top can be kept in its decay products. Therefore one can use the decay products to reconstruct the top-quark polarization information,
and measure $\tan{\beta}$ with better precision.

If charged Higgs boson is heavier than top quark, it can be mainly produced in three ways: (1) $pp\rightarrow \gamma/ Z\rightarrow H^+H^-$; (2) $gb\rightarrow tH^-(\bar{b}g\rightarrow \bar{t}H^+) $; and (3) $q\bar{q}^{\prime}\rightarrow W\rightarrow AH^{\pm}/hH^{\pm}/{HH^{\pm}}$. The cross section of process (1) decreases with $m_{H^{\pm}}$ more rapidly compared to the other two, and the process (3) contains other unknown parameter $m_A$. In order to measure $\tan\beta$ in good accuracy,
we adopt $tH^-/\bar{t}H^+$ associated production

\begin{equation}
gb\rightarrow tH^-\rightarrow t\bar{t}b
\label{signal_1}
\end{equation}
\begin{equation}
g\bar{b}\rightarrow \bar{t}H^+\rightarrow \bar{t}t\bar{b}
\label{signal_2}
\end{equation}

These two processes generate the same final states: one top-quark, one anti-top-quark and a bottom-quark. If we want to probe the $H^+\bar{t}b$ coupling, we should take care of the different origin of the anti-top (the same if taking the top quark) in two processes: the $\bar{t}$ of process~(\ref{signal_1}) is the decay product of charged Higgs boson, while the $\bar{t}$ in process~(\ref{signal_2}) is associated produced with the charged Higgs boson. We calculate the helicity amplitudes of different processes and obtain their degree of top polarization respectively as \cite{Cao:2013ud},
\begin{equation}
D_{\rm decay}\equiv\frac{\Gamma(\bar{t}_L) - \Gamma(\bar{t}_R)}{\Gamma(\bar{t}_L) + \Gamma(\bar{t}_R)} = \frac{(m_t \cot\beta)^2 - (m_b\tan\beta)^2}{(m_t\cot\beta)^2 + (m_b\tan\beta)^2}
\end{equation}
\begin{equation}
D_{\rm prod}(\hat{s}) \equiv \frac{\hat{\sigma}(t_R)-\hat{\sigma}(t_L)}{\hat{\sigma}(t_R)+\hat{\sigma}(t_L)} = \frac{(m_t\cot\beta)^2-(m_b \tan\beta)^2}{(m_t\cot\beta)^2+ (m_b \tan\beta)^2 } \times \hat{R}_{\rm prod}.
\label{dilution_def}
\end{equation}

\begin{figure}[bht]
\begin{center}
\includegraphics[width=\linewidth]{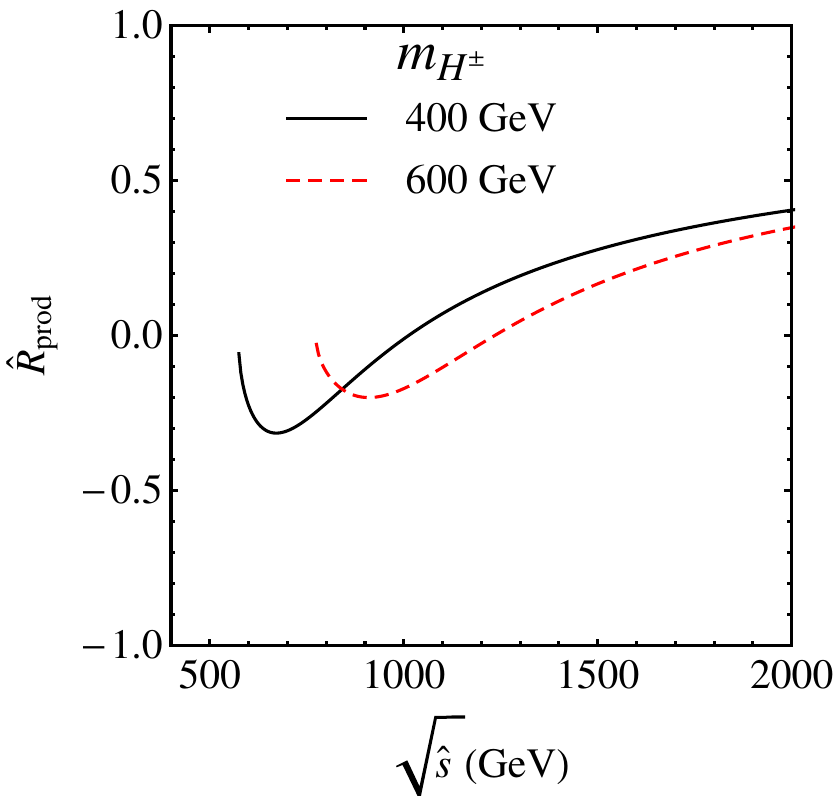}
\end{center}
\caption{\it The dilution factor in Eq.~\ref{dilution_def} as a function of the energy of the overall c.m. frame ($\sqrt{\hat{s}}$) with $m_{H^\pm}=400\rm {~ GeV}$(solid black) and $m_{H^\pm}=600\rm{~GeV}$ (dashed red){\cite{Cao:2013ud}}. }\label{fig:Rprod}
\end{figure}

The dilution factor $\hat{R}_{\rm prod}$ as a function of the c.m. energy $\sqrt{\hat{s}}$ of the sub-process for $m_{H^{\pm}}=400$ GeV and  $m_{H^{\pm}}=600$ GeV is shown in Fig.~\ref{fig:Rprod}. From the curves, we can find the absolute value of $\hat{R}_{\rm prod}$ is less than 0.5 and the sign of the dilution factor varies from negative in the threshold region of the $tH^-$ to positive in the large invariant mass region. The sign of the dilution factor is very important for the measurement of top polarization. Therefore, we focus our attention on the anti-top from the $H^-$ decay rather than the associated production with the $H^+$. It means that the anti-top production with charged Higgs boson
should be treated as background if we wish to measure the charged Higgs boson interaction utilizing top polarization information.

\section{Discovering charged Higgs boson at the LHC}

In order to measure $\tan\beta$, the first task is to discover the charged Higgs boson, i.e. to suppress the backgrounds.  The task can be divided into two: one is to distinguish the $gb\rightarrow tH^-$ from $g\bar{b}\rightarrow \bar{t}H^+$, and the other is to suppress the
other SM backgrounds. We require $\bar{t}\rightarrow l^-\bar{b}\bar{\nu}$ and $t\rightarrow bjj$. The anti-top polarization can be
inferred via the angular distribution of charged leptons as described below. The signal is
\begin{equation}
gb \rightarrow tH^-\rightarrow (W^+b)(\bar{t}b) \rightarrow (jjb)(l^-\bar{\nu}\bar{b}b)
\end{equation}
Irreducible backgrounds are
\begin{equation}
g\bar{b} \rightarrow \bar{t}H^+\rightarrow (W^-\bar{b})(t\bar{b}) \rightarrow (l^-\bar{\nu}\bar{b})(jjb\bar{b})
\end{equation}
The SM
backgrounds are
\begin{eqnarray}
&pp\rightarrow t\bar{t}j_b\rightarrow bW^+\bar{b}W^-j_b\rightarrow j_bj_bj_bjjl^-\bar{\nu} \nonumber \\
&pp\rightarrow t\bar{t}j\rightarrow bW^+\bar{b}W^-j\rightarrow j_bj_bjjjl^-\bar{\nu},
\end{eqnarray}
where $j_b$ means b-jet containing one b or $\bar b$. Note that all of our analysis are based on parton level and
leading order estimation. There are five jets in the final states. We order the five jets by their $p_T$ and examine their $p_T$ distributions~\cite{Cao:2013ud}. Based on these information we can introduce the basic $p_T$ cut on signal and backgrounds as
\begin{equation}
p_T(j_{\rm 1st}) \geq 120~{\rm GeV}, p_T(j_{\rm 2nd})\geq 80~{\rm GeV}, p_T(j_{\rm 3rd})>60~{\rm GeV};
\label{hardcut}
\end{equation}
In order to suppress the huge SM backgrounds, we need to reconstruct the intermediate states and isolate the extra jet. The detailed analysis can be
 found in  \cite{Cao:2013ud}.
In order to distinguish $gb\rightarrow tH^-$ from $g\bar{b}\rightarrow \bar{t}H^+$, we apply the extra cut as
\begin{equation}
\Delta M_{\bar{t}j_{\rm extra}}\equiv \left| M_{\bar{t} j_{\rm extra}}-M(H^\pm) \right| \leq 5{\rm ~GeV}, p_T(j_{\rm extra}) \geq 120~{\rm GeV}.
\end{equation}

\begin{table}
\begin{center}
\caption{Number of events of SM backgrounds at the 14 TeV LHC with an integrated luminosity of $100~{\rm fb}^{-1}$. This table is from Ref. \cite{Cao:2013ud}. }
\label{tbl-efficiency}
\begin{tabular}{c|c|c}
\hline
$\tan\beta$ &  \multicolumn{2}{c}{SM backgrounds}\tabularnewline
\hline
 &  $t\bar{t}j$ & $t\bar{t}b$\tabularnewline
\hline
\hline
 &  $1.075\times10^{7}$ & 234000\tabularnewline
$p_T$ cuts &  $2.12\times10^{6}$ & 25052\tabularnewline
$\Delta M_{\bar{t}j_{\rm extra}}$ & 39238 & 386\tabularnewline
$p_{T}(j_{{\rm extra}})$ & 14942 & 171\tabularnewline
$b$ tagging &  299 & 102\tabularnewline
\hline
\hline
Event No. & \multicolumn{2}{c}{401}\tabularnewline
\hline
\end{tabular}
\end{center}
\end{table}

Finally, in order to suppress the backgrounds further, we demand the extra jet to be a $b$ jet and choose the b-tag efficiency as 60\% and mis-tagging efficiency as 2\%.
The cut efficiency for backgrounds/signal is listed in Tab. \ref{tbl-efficiency}, Tab. \ref{tbl-efficiency1}and Tab. \ref{tbl-efficiency2}.
From the tables the cut conditions can suppress the SM background to more than 4 orders of magnitude and irreducible backgrounds to more than 2 orders of magnitude, while
the signals have enough event numbers.
We can get the significance of the signal over backgrounds well above $5\sigma$ for a broad range of $\tan{\beta}$.
Even for hardest case with $\tan{\beta}=6$, there are more than 300 events survive for $m_{H^\pm}=400$ GeV at the 14 TeV LHC with an integrated luminosity of 100 $fb^{-1}$.

\begin{table}
\begin{center}
\caption{Number of events of the signal and irreducible backgrounds at the 14 TeV LHC with an integrated luminosity of $100~{\rm fb}^{-1}$ for $m_{H^\pm}=400~{\rm GeV}$ and  $\tan\beta=1, 6$. This table is from Ref. \cite{Cao:2013ud}. }
\label{tbl-efficiency1}
\begin{tabular}{c||c|c|c|c}
\hline
$\tan\beta$ & \multicolumn{2}{c|}{1} & \multicolumn{2}{c}{6}  \tabularnewline
\hline
 & $tH^{-}$ & $\bar{t}H^{+}$ & $tH^{-}$ & $\bar{t}H^{+}$  \tabularnewline
\hline
\hline
 & 23310 & 23300 & 1255 & 1227  \tabularnewline
$p_T$ cuts & 11843 & 13466 & 687 & 719  \tabularnewline
$\Delta M_{\bar{t}j_{\rm extra}}$ & 4980 & 368 & 672 & 20  \tabularnewline
$p_{T}(j_{{\rm extra}})$ & 3910 & 305 & 532 & 16  \tabularnewline
$b$ tagging & 2346 & 183 & 312 & 10 \tabularnewline
\hline
\hline
Event No. & \multicolumn{2}{c|}{2529} & \multicolumn{2}{c}{322}  \tabularnewline
$S/B$ & \multicolumn{2}{c|}{6.3} & \multicolumn{2}{c}{0.8}  \tabularnewline
$S/\sqrt{B}$ & \multicolumn{2}{c|}{126.3} & \multicolumn{2}{c}{16.1}  \tabularnewline
$S/\sqrt{S+B}$ & \multicolumn{2}{c|}{54.1} & \multicolumn{2}{c}{26.9}  \tabularnewline
\hline
\end{tabular}
\end{center}
\end{table}

\begin{table}
\begin{center}
\caption{Number of events of the signal and irreducible backgrounds at the 14 TeV LHC with an integrated luminosity of $100~{\rm fb}^{-1}$ for $m_{H^\pm}=400~{\rm GeV}$ and $\tan\beta=40$. This table is from Ref. \cite{Cao:2013ud}. }
\label{tbl-efficiency2}
\begin{tabular}{c||c|c}
\hline
$\tan\beta$ &  \multicolumn{2}{c}{40} \tabularnewline
\hline
 &   $tH^{-}$ & $\bar{t}H^{+}$ \tabularnewline
\hline
\hline
 &  24660 & 23520 \tabularnewline
$p_T$ cuts & 14421 & 13890 \tabularnewline
$\Delta M_{\bar{t}j_{\rm extra}}$ &  5680 & 383 \tabularnewline
$p_{T}(j_{{\rm extra}})$ & 4375 & 310 \tabularnewline
$b$ tagging & 2625 & 186 \tabularnewline
\hline
\hline
Event No. &  \multicolumn{2}{c}{2811} \tabularnewline
$S/B$ &  \multicolumn{2}{c}{7.0} \tabularnewline
$S/\sqrt{B}$ & \multicolumn{2}{c}{140.3} \tabularnewline
$S/\sqrt{S+B}$ &  \multicolumn{2}{c}{56.7} \tabularnewline
\hline
\end{tabular}
\end{center}
\end{table}

\section{Measuring $\tan\beta$ at the LHC utilizing top polarization}

With the reconstructed $\bar{t}$, $j_{extra}$ and $H^-$, we can measure the anti-top's polarization by angular distribution of top decay products. As usual, we define an angle between the charged lepton momentum in the rest frame of $\bar{t}$ to the anti-top momentum in the rest frame of $H^-$
\begin{equation}
\frac{d\sigma}{\sigma d\cos\theta_{\rm hel}}=\frac{1}{2} (1+D\cos\theta_{\rm hel}).
\end{equation}
If we can measure the distribution of $\frac{d\sigma}{\sigma d\cos\theta_{\rm hel}}$, we can get the polarization of the anti-top quark by
 \begin{equation}
D=3\sum_{i=1}^{10} \cos\theta_i \left(\frac{d\sigma}{\sigma d\cos\theta}\right)_i\Delta \cos\theta
 =\frac{3\sum_{i=1}^{10} \cos\theta_i N_i}{\sum_{i=1}^{10} N_i},
\label{d-definition}
\end{equation}
where $N_i$ means the rescaled event number of the ith bin in the distribution of $\frac{d\sigma}{\sigma d\cos\theta_{\rm hel}}$. In our analysis, there are only 10 bins distributed between $\cos\theta_{\rm{hel}}=-1$ and $1$, so the i varies from 1 to 10. For simplicity, we introduce the statistical error of the degree of anti-top polarization as following:
\begin{equation}
\Delta D=\sqrt{\sum_{i=1}^n\left|\frac{\partial D}{\partial N_i}\right|^2 \left({\Delta N_i}\right)^2}.
\end{equation}

\begin{figure}[bht]
\begin{center}
 \includegraphics[width=0.4\textwidth]{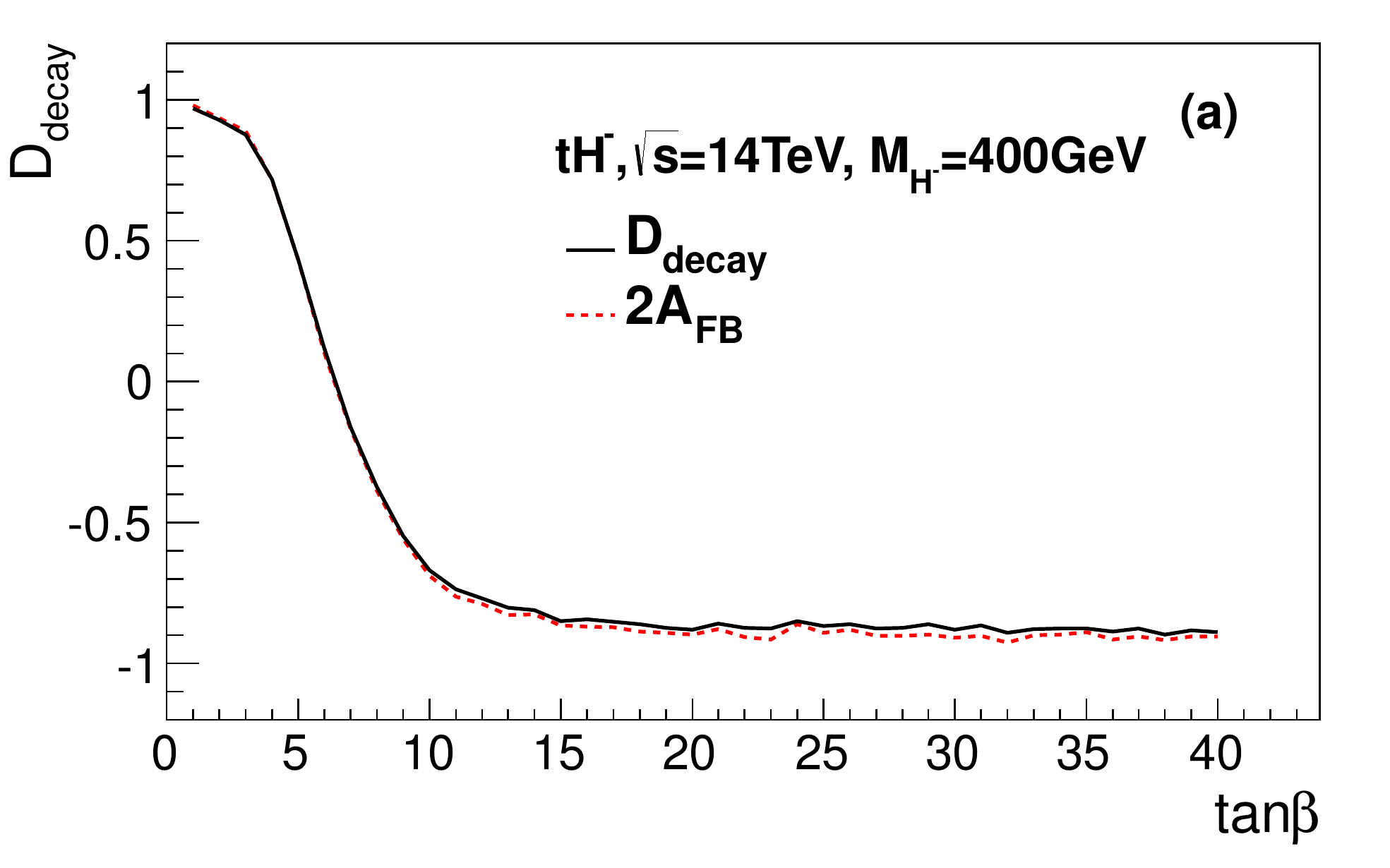}
  \includegraphics[width=0.4\textwidth]{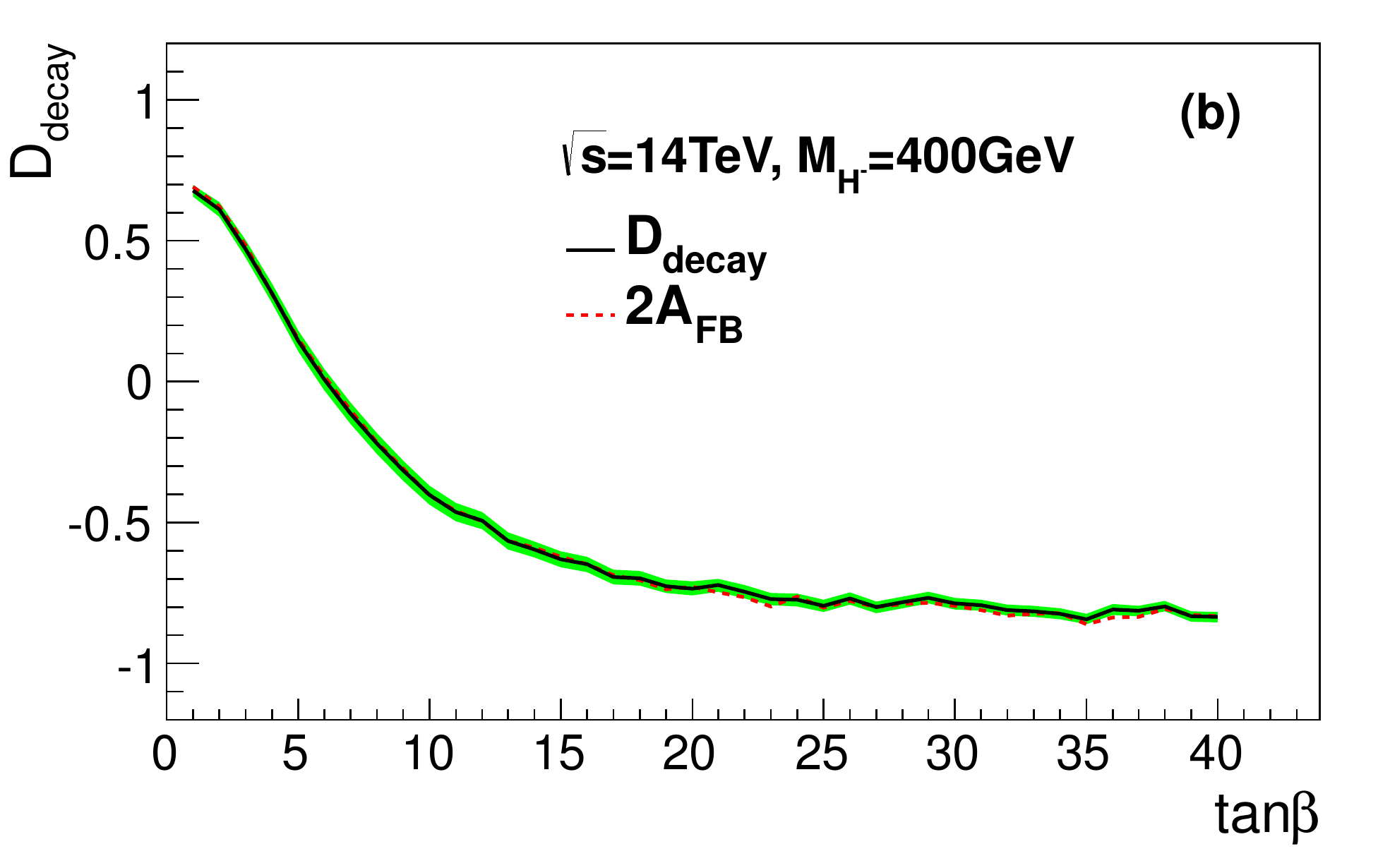}
  \end{center}
  \caption{\it (a) The degree of polarization of the anti-top quark as a function of $\tan\beta$ of the $tH^-$ signal event and (b) of all the signal and background processes with $m_{H^\pm}=400~\rm{ GeV}$. The solid black curve shows the degree of polarization defined in Eq.~\eqref{d-definition}; the dashed red curve shows $2A_{FB}$. The green band in (b) represents only the statistical uncertainties.}\label{fig:toppol}
\end{figure}

The polarization degree of the anti-top quark as a function of $\tan{\beta}$ is shown in Fig.~\ref{fig:toppol} with equation \ref{d-definition}. $A_{FB}\equiv\frac{\sigma_F -\sigma_B}{\sigma_F + \sigma_B}$ is also plotted for comparison. From the figures, we can see that the anti-top quark polarization is a good probe for a wide range for $\tan{\beta}$,
especially for the intermediate $\tan{\beta}$. For $\tan]beta$ in this region, it is hard to measure by using the cross section information. However $D_{\rm decay}$ changes rapidly in the region of $\tan{\beta}=5\sim10$. This feature helps us to determine $\tan{\beta}$ with better accuracy. Figure~\ref{fig:toppol}(b) tell us that the polarization can not reach $\pm1$ because of the SM backgrounds and the signal events loss by the cut conditions.

\section{Conclusion and discussion}

In this paper, we discussed briefly the motivations for charged Higgs boson. Among them we focused on the role of charged Higgs boson to reveal
the possible new paradigm, which is the new way to understand the electro-weak symmetry breaking and CP violation. In practice, discovering charged
Higgs boson and measuring its properties, like $\tan\beta$, are crucial to distinguish various models.

We wish to emphasize that the charged Higgs boson is likely only one piece of big puzzle, though the quite important piece.



\subsection*{Acknowledgment}

 This work was supported in part by the Natural Science Foundation
 of China (Nos. 11135003 and 11375014).

\nocite{*}
\bibliographystyle{elsarticle-num}
\bibliography{martin}






\end{document}